%% file: main.tex
\documentclass[letterpaper,twocolumn,10pt]{article}
\usepackage{usenix2019_v3}

\usepackage{xspace}
\usepackage{amsmath}
\usepackage{graphicx}
\usepackage[colorinlistoftodos]{todonotes}
\usepackage{subcaption}
\usepackage{booktabs}
\usepackage{authblk}
\usepackage{soul}
\usepackage{algorithm}

\usepackage[noend]{algpseudocode}

\title{\toolname: Defending Against Package Typosquatting}
\author[1]{Matthew Taylor}
\author[1]{Ruturaj K. Vaidya}
\author[1]{Drew Davidson}
\author[2]{Lorenzo De Carli}
\author[3]{Vaibhav Rastogi}
\affil[1]{University of Kansas}
\affil[2]{Worcester Polytechnic Institute}
\affil[3]{Google}
\date{}

\include{macros}
\begin{document}
\maketitle
\input{abstract}
\input{introduction}
\input{background}
\input{overview}
\input{evaluation}
\input{discussion}
\input{related_work}
\input{conclusion}

\bibliographystyle{plain}
\bibliography{bibliography}
\end{document}

%% file: macros.tex
\newcommand{\myparagraph}[1]{\vspace{0.05in}\noindent\textbf{#1}:}

\newcommand{\toolname}{SpellBound}
\newcommand{\tooltext}{\toolname\xspace}

%% file: abstract.tex
\begin{abstract}

  Package managers for software repositories based on a single programming language are very common. Examples include npm (JavaScript), and PyPI (Python). These tools encourage code reuse, making it trivial for developers to import external packages. Unfortunately, repositories' size and the ease with which packages can be published facilitates the practice of \textit{typosquatting}: the uploading of a package with name similar to that of a highly popular package, typically with the aim of capturing some of the popular package's installs. Typosquatting has serious negative implications, resulting in developers importing malicious packages, or---as we show---code clones which do not incorporate recent security updates.

  In order to tackle this problem, we present \toolname{}, a tool for identifying and reporting potentially erroneous imports to developers. \toolname{} implements a novel typosquatting detection technique, based on an in-depth analysis of npm and PyPI. Our technique leverages a model of lexical similarity between names, and further incorporates the notion of package popularity. This approach flags cases where unknown/scarcely used packages would be installed in place of popular ones with similar names, before installation occurs. We evaluated \toolname{} on both npm and PyPI, with encouraging results: \toolname{} flags typosquatting cases while generating limited warnings (0.5\% of total package installs), and low overhead (only 2.5\% of package install time). Furthermore, \toolname{} allowed us to confirm known cases of typosquatting and discover one high-profile, unknown case of typosquatting that resulted in a package takedown by the npm security team.

\end{abstract}

%% file: introduction.tex
\section{Introduction}
\label{sec:Introduction}

%Catastrophic for the installer and the client of the project
\textit{Package managers} are tools which automate the complex task of deploying 3rd-party dependencies into a codebase, abstracting away the provenance of the dependency; when the user invokes a command to install the package by name, the given package will be downloaded from a remote repository, alongside the full set of additional packages upon which it transitively depends.  One of the most common uses of package managers is in the context of large repositories of code packages based on a single programming language.
Package managers are undeniably useful, with open, free-for-all repositories like npm for Node.js/JavaScript, PyPI for Python, the NuGet Gallery for Microsoft's .NET Framework, and crates.io for Rust, collectively serving billions of packages per week. Despite their utility, package managers also come with problems.

The ease with which code can be imported facilitates incorrect imports. Installing an unintended code dependency can be catastrophic, but happens as easily as mistyping a single character on the command line.
Furthermore, the open, uncurated nature of these repositories means that any developer can upload a package with a name of  their choosing and it will be treated with equal trust as any other package in the repository. This circumstance gives rise to \textit{typosquatting}, whereby a developer uploads a ``perpetrator'' package that is \textit{confusable} with an existing ``target'' package due to name similarity.

The process by which typosquatting acts is simple: the user, intending to install the target package, accidentally requests the name of the confusable perpetrator package.
Determining why perpetrators packages are created and uploaded is a challenging and ill-defined problem, as solving it requires inferring the intent of the package author. The perpetrator may wish to intentionally confuse users into installing a malicious payload, seek to increase the visibility of their own benign code, or may have created a confusable name by happenstance, without realizing it. A typosquatting perpetrator might even upload a placeholder package to prevent an attacker from leveraging the given name.
Regardless of the intent, the result is the same: users are confused into importing the incorrect package into their code.

Typosquatting has numerous detriments, both to developers who integrate a perpetrator package into their codebase, and to the end-users of such a codebase. An overtly malicious perpetrator may include Trojan functionality that attacks the client when run~\cite{npm_bitcoin_miner, npm_password_cvc}. Additionally, many package managers invoke configuration hooks bundled with the package at install time, often manifested as shell scripts that run with the privileges of the user. 
Multiple packages that open reverse shells when installed have been removed from npm~\cite{npm_system_wipe, npm_preinstall_ssh, npm_crypto_hijacking}. 
%Theoretically, a perpetrator can also exhibit purely-benign functionality until it has escaped scrutiny, at which time it can be updated with malicious functionality.
Even in cases where the perpetrator package is not overtly malicious, it can confuse the user and weaken the integrity of the system. Ironically, a perpetrator might clone a victim to keep it out of the hands of an attacker but allow the clone to fall behind as the target is updated, exposing users of the clone to latent vulnerabilities that have been patched out of the target.

Typosquatting is a difficult problem to detect manually, as it confuses manual inspection by definition. In this work, we develop \toolname{}, a novel typosquatting detection technique to discover and prevent incidents of typosquatting before they can damage the user. \toolname{} can be used to detect typosquatting incidents before they happen, or to detect possible perpetrator packages within a package repository.

To illustrate typosquatting, and the benefits of our approach, consider the example of \textit{loadsh}, an npm package that \toolname{} reported to be typosquatting the popular \textit{lodash} package. Because \textit{loadsh} is a transposition of the ``a'' and ``d'' characters of \textit{lodash}, our techniques detected that the package names are easily confusable. We confirmed that \textit{loadsh} was being used uninentitionally by emailing the maintainers of packages that used \textit{loadsh}. Three \textit{loadsh}-dependant package maintainers responded to our email, all of whom acknowledged that they had intended to install \textit{lodash} and indicated that they would change their dependency. Many of the packages using \textit{loadsh}, including those maintained by our respondents, had been victims for over a year.

The \textit{loadsh} incident exemplifies several stealthy aspects of package name typosquatting; not only are the developers who use \textit{loadsh} victims of a typosquatting attack, so are packages that \textit{transitively} depends on \textit{loadsh} (i.e. those codebases that depend on a package that accidentally uses \textit{loadsh}). Thus, it is possible to be a victim to a typosquatting attack without personally making a typo.
Another difficulty in detecting packages like \textit{loadsh} is that they may not exhibit malicious behavior.
Indeed, \textit{loadsh} does not include any malicious functionality - the perpetrator package is an exact snapshot copy of \textit{lodash} version 4.17.11, the current version of the target at the time at which the typosquatting package was created (the target package \textit{lodash} is currently at version 4.17.15).
Nevertheless, the perpetrator still has a negative impact; because the perpetrator package has not been updated, its victims were effectively using an outdated version of \textit{lodash}.
In the case of this example, the older version has been reported to contain prototype pollution vulnerabilities~\cite{lodash_prototype_pollution}, effectively leaving victims of \textit{loadsh} open to attacks that have already been patched in the current version of \textit{lodash}.
When \textit{loadsh} was reported, 63 other package depended on it. Each of these dependents were, by extension, vulnerable to prototype pollution.
After \toolname{} reported \textit{loadsh} as a typosquatting perpetrator, we contacted the npm security team, who verified our results, deprecated \textit{loadsh}, and took over ownership of the package.

% explain capabilities
As indicated by the \textit{loadsh} example above \toolname{} can be used to detect if a given package is a perpetrator of a typosquatting attack. At high level, \toolname{} intercepts and analyzes package install requests. First, it checks if a given package is not popular. If so, it checks if the given package's name is \textit{lexically similar} to that of a popular one (we describe and motivate our notions of popularity and similarity in Section~\ref{sec:Overview}). If both conditions are met, \toolname{} concludes that the user is at risk of installing a typosquatting perpetrator, and issues an alert before the package is fetched. Furthermore, it presents a suggestion for the likely correct package name that is being typosquatted.
%when the package manager command is issued. When \toolname{} detects that the user is at risk of getting a typosquatting perpetrator, it will issue an alert before the package is fetched, alongside a suggestion for what the likely target package is that is being typosquatted. 

%Basic steps have been taken to mitigate these typosquatting,
%though the issue persists. In this paper, we present \toolname{}: a robust and 
%extensible modification to the package manager front-end designed to protect 
%against a variety of typosquatting attacks. By performing specific %transformations on a given package name, 

Overall, our work makes the following contributions:
\begin{itemize}
    \item We highlight the security implications of typosquatting.
    \item We study the extent to which typosquatting exists in npm and PyPI.
    \item We present \toolname{}, an enhancement to the package manager front-end which protects users against typosquatting attacks.
    \item We evaluate the efficacy of \toolname{}. We show it offers a higher level of security while incurring a 2.5\% overhead during package installation. Additionally, we demonstrate that \toolname{} is non-intrusive, as it affects less than 1\% of all weekly downloads for popular package repositories.    
\end{itemize}

The rest of this paper is structured as follows: Section~\ref{sec:Background} provides background on package managers and typosquatting attacks. Section~\ref{sec:Overview} describes and motivates the design of \toolname{}. Section~\ref{sec:Evaluation} evaluates \toolname{}'s performance. Section~\ref{sec:Discussion} discusses limitations and possible extensions of our work. Section~\ref{sec:related_work} examines the related work. Finally, Section~\ref{sec:Conclusion} concludes the paper.

%%% Local Variables:
%%% mode: latex
%%% TeX-master: t
%%% End:

%% file: background.tex
\section{Background}
\label{sec:Background}

In this section, we give background information necessary to understand the need for a tool like \toolname. In particular, we show how the current landscape of package management enables typosquatting, and describe previous attacks that use typosquatting to deliver malicious payloads. We discuss the context that makes typosquatting a pernicious problem for many of the repository stakeholders, including end-users of applications, application developers, package providers, and the maintainers of repositories themselves.

\subsection{Package Repositories}

\begin{table}[]
\begin{tabular}{lll}
\toprule
                          & npm            & PyPI        \\
\midrule
Packages                  & 1,221,705      & 221,041     \\
Weekly Downloads          & 17,872,179,641 & 997,624,343 \\
Avg. Dependency Tree Size & 57.27          & 4.58       
\end{tabular}
\vspace{0.5cm}
\caption{Usage statistics for npm and PyPI: Both repositories serve significant numbers of highly interdependent packages on a weekly basis.}
\label{tab:repo_stats}
\end{table}

The use of package repositories for managing dependencies is incredibly popular. They simplify the use of third-party code, which in turn has obvious benefits. It encourages code reuse; it allows expertly-written and well-vetted codebases to be deployed by more developers; and it leverages the knowledge of the broader software development community even for highly-custom projects. 
For these reasons, successful repositories may grow to enormous size.
The first two rows of Table~\ref{tab:repo_stats} show the current size and weekly download counts for npm and PyPI, as reported by the repository maintainers. As the table shows, they contain hundreds of thousands (in the case of PyPI), or even millions (in the case of npm) of publicly available packages.
The total number of weekly downloads served are nearly 1 billion in the case of PyPI and over 17 billion in the case of npm. 
%Additionally, dependencies comprise a significant portion of application code: a recent report by the software security company Contrast Security found that 79\% of application code came from third parties~\cite{contrast}. Given the bulk of code existing in dependencies, it is infeasible to expect developers to manually inspect every line of code that they integrate into their project, even in cases when that code is human readable.

Much of the complexity of package management is due to the interdependence of packages. For example, the popular npm package \textit{webpack-dev-server} (6.6 million weekly downloads) declares 33 dependencies of its own. These 33 dependencies require further packages to be installed (the transitive dependencies of \textit{webpack-dev-server}). 
In total, \textit{webpack-dev-server} has 391 transitive dependencies. 
Running \textit{webpack-dev-server} requires that all 391 packages are installed. 
Furthermore, these packages span many distinct development teams, each of which may update out of step with one another, introducing new functionality and behavior. This is in line with the  general trend of code reuse in software development: a recent report by the software security company Contrast Security found that 79\% of application code came from third parties~\cite{contrast}. Given the bulk of code existing in dependencies, it is infeasible to expect developers to manually vet every package or piece of code that they integrate into their project.

% Manually navigating these vast and complex package repositories to find and %install dependencies is a tedious process. Package managers automate this %process while abstracting it from the user. With a single command, %developers can leverage package managers to download a large number of %dependencies in a matter of seconds. It's not uncommon for packages to rely %on dozens, or even hundreds, of other packages~\cite{zimmermann_small_2019}.
%\subsection{Package Managers}
Package manager frontends automate the complex and tedious task of fetching, configuring, and updating a package and its transitive dependencies.
When a user issues a command like \texttt{npm install webpack-dev-server}, the frontend relies on the package's metadata to build a spanning tree of the package dependency graph (referred to internally as the package dependency tree), and then installs each package node in the tree. Similarly, the command \texttt{npm update} updates the package dependency tree for the current set of packages, and ensures that the most recent compatible versions of dependencies are deployed. The third row of Table~\ref{tab:repo_stats} shows the average size of the dependency tree for the two package managers we study. It is notable that there is significant interdependence among packages. 

While package managers save users a significant amount of time, they do not help with the herculean task of vetting imported code; if anything, they complicate it. The key design goal of package manager frontends is that they make fulfilling dependencies opaque to the user. 
As a result, the provenance of a package is also obscured - a user need not explicitly trust the developer of a package they (transitively) use, nor even know \textit{who} uploaded the code to the package repository.
Once a package is registered to the repository, it is given equal trust as any other package on the repository, and may be freely integrated into applications or other packages.

%Treating all packages in the repository equally has a key utility benefit: 
%Numerous distinct developer teams can rely on each other with no additional burden to the user and little explicit collaboration between the various package providers.

% \begin{figure}[t]
% \includegraphics[clip, trim=4.9cm 4.5cm 4.6cm 4.5cm, width=\linewidth]{images/npm_download_distribution.pdf}
% \caption{The download distribution for packages on npm.}
% \label{fig:npm_download_distribution}
% \end{figure}

% \begin{figure*}[t]
%     \centering
%     \begin{subfigure}[b]{0.5\textwidth}
%     %\fbox{
%     \includegraphics[clip, trim=4.9cm 4.5cm 4.6cm 4.5cm, width=\linewidth]{images/npm_download_distribution.pdf}
%     %}
%         \caption{The download distribution for packages on PyPI}
%         \label{fig:npm_download_distribution}
%     \end{subfigure}
%     ~ %add desired spacing between images, e. g. ~, \quad, \qquad, \hfill etc. 
%       %(or a blank line to force the subfigure onto a new line)
%     \begin{subfigure}[b]{0.5\textwidth}
%     %\fbox{
%         \includegraphics[clip, trim=4.9cm 4.5cm 4.6cm 4.5cm, width=\linewidth]{images/pypi_download_distribution.pdf}
%      %   }
%         \caption{The download distribution for packages on npm.}
%         \label{fig:pypi_download_distribution}
%     \end{subfigure}
%     ~ %add desired spacing between images, e. g. ~, \quad, \qquad, \hfill etc.
%     \caption{Pictures of animals}\label{fig:animals}
% \end{figure*}

\begin{figure}[t]
\includegraphics[clip, trim=4.9cm 4.5cm 4.6cm 4.5cm, width=\linewidth]{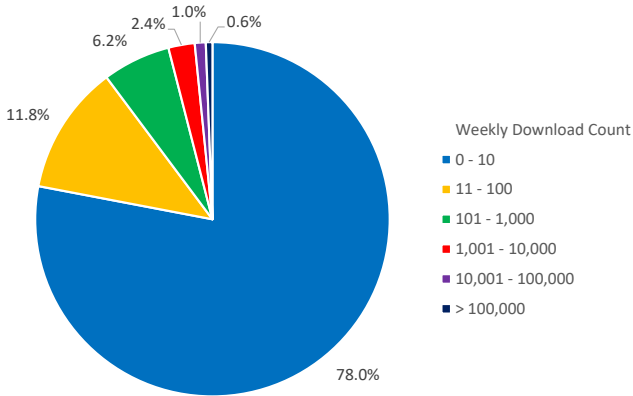}
\caption{Download distribution for packages on npm.}
\label{fig:npm_download_distribution}
\vspace{-0.5cm}
\end{figure}

\begin{figure}[t]
\includegraphics[clip, trim=4.9cm 4.5cm 4.6cm 4.5cm, width=\linewidth]{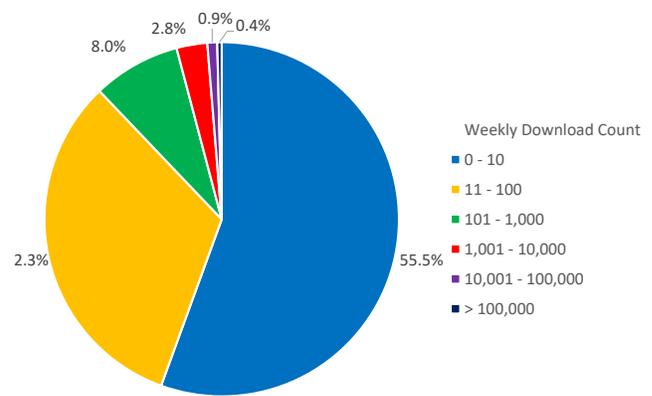}
\caption{Download distribution for packages on PyPI.}
\label{fig:pypi_download_distribution}
\vspace{-0.35cm}
\end{figure}

\myparagraph{Characterization of Package Downloads}
The majority of package downloads are due to a small number of packages.
Based on the self-reported repository download counts, we classified the popularity of  packages across npm and PyPI. 
Figure~\ref{fig:npm_download_distribution} and Figure~\ref{fig:pypi_download_distribution} show the distribution of downloads across npm and PyPI, respectively. 
A majority of the packages for both repositories are downloaded between zero and ten times per week. Only a small fraction of packages see a high degree of popularity. However, the packages composing the smallest portion of each figure actually receive more downloads than the packages in all remaining portions combined. Locating desired packages in this ocean of unpopular ones without assistance can be challenging.

\subsection{Factors Contributing to Typosquatting}
The automated nature of package managers has enormous utility. However, this also enables misuse, namely typosquatting. 
We propose that the following aspects of package repositories contribute to the threat of typosquatting:
\begin{itemize}
    \item The open-source nature of repositories means that any user can upload a package, and it will be given equal trust with any other package.
    \item The provenance of a package is opaque to the user, and the interdependence between packages makes their behavior difficult to vet manually.
    \item The distribution of packages means there are a small number of ``juicy'' typosquatting targets, and a large number of package from which a typosquatting attack could be launched.
\end{itemize}

We now review select cases of historical typosquatting, and describe the challenges in detecting typosquatting reliably.
%To support our claim that package repositoriesWe now describe how these factors have contributed to past incidents of typosquatting against the package managers we study.

%Three enabling aspects of typosquatting: 
% - anybody can upload a package
% - all packages are treated equally, even though some are super-popular
% - the provenance of packages are opaque

\subsection{Historical Package Typosquatting}

The degree to which typosquatting has historically occurred is difficult to capture, due in part to the highly subjective nature of what constitutes typosquatting. Indeed, there exist cases of package name similarity where intent may appear benign or ambiguous. In practice, most packages that are flagged by repositories exhibit overtly malicious functionality, and are retroactively deemed typosquatting perpetrators by a qualitative manual analysis. It is also important to observe that not all malicious packages perform typosquatting.

As an example of the complexities of determining typosquatting and its intent, consider the \textit{js-sha3} typosquatting campaign. On October 25th, 2019, 25 packages were simultaneously identified by Microsoft Vulnerability Research and taken down by the npm security team: 
\textit{
zs-sha3,
ns-sha3,
ks-sha3,
jw-sha3,
jsmsha3,
js-wha3,
js-sxa3,
js-sla3,
js-sja3,
js-sia3,
js-shq3,
js-she3,
js-shc3,
js-shas,
js-sha7,
js-rha3,
js-qha3,
js-cha3,
js-3ha3,
jr-sha3,
jq-sha3,
jc-sha3,
j3-sha3,
hs-sha3,
bs-sha3.
}

Upon close inspection, all those packages were determined to have malicious intent, and all package names were close, according to Levenshtein distance, to the victim package \textit{js-sha3}. However, not all packages names were likely to confuse the user. For example, \textit{js-sxa3} requires replacing the ``h'' with an ``x''. It is unlikely that a developer would misremember \textit{js-sha3} as \textit{js-sxa3} (the package being an implementation of the SHA-3 algorithm). A typo is equally unlikely on a QWERTY keyboard, given the distance between ``h'' and ``x''. As discussed in Section~\ref{sec:TyposquattingSignals}, we take the stance of only flagging cases where there is strong likelihood that name similarity may confuse the user. While this causes us to ignore some cases (as \textit{js-sxa3} above), it has the advantage to avoid generating an excessive number of warnings.

One may also be tempted to solve these ambiguities by always attempting to identify malicious intent, regardless of whether typosquatting occurs. In practice, this is challenging and currently impossible to achieve reliably. Source code analyses might be employed to catch overtly malicious behavior in a perpetrator package, but they have difficulty detecting obfuscated payloads.  JavaScript is a particularly difficult target to analyze - recent work has shown that JavaScript can be automatically obfuscated to appear syntactically indistinguishable from benign code to modern detectors~\cite{fass_hidenoseek:_2019}.  Furthermore, the highly dynamic nature of JavaScript means that malicious functionality may not appear until the script is deployed.
%A further challenge in using source code analysis is that the payload may be grayware. In general, using source code to infer the intent of the package uploader (and potentially matching that to the intent of the package downloader) remains an open problem.

%Detecting typosquatting packages is challenging. Source code analyses might be employed to catch overtly malicious behavior in a perpetrator package, but may have difficulty detecting obfuscated payloads. Automatic detection of malicious code remains a challenging open problem. Javascript is a particularly difficult target to analyze - recent work has shown that Javascript can be automatically obfuscated to appear syntactically indistinguishable from benign code to modern detectors~\cite{hidenoseek}.  Furthermore, the highly dynamic nature of Javascript means that malicious functionality may not appear until the script is deployed.
%A further challenge in using source code analysis is that the payload may be grayware.
%Using source code to infer the intent of the package uploader, and matching that to the intent of the package downloader, remains another open problem. 

Currently, the standard technique for removing these packages is manual and reactive. Users who believe a package is performing malicious typosquatting can file a report to the repository maintainers, who will then investigate the claim. Should the maintainers agree with the reporter, the package will be removed. This approach does little to prevent the installation of malicious packages and fails to protect users from the consequences.

Despite the shortcomings of this approach, hundreds of package takedowns have been issued that involve package names similar to a popular target. We believe this number to be a lower bound on the total number of typosquatting attempts. 
Due to the ease with which packages can be registered to a repository, the differential of effort favors the attacker; the 25 packages reported by Microsoft above exhibit many of the hallmarks of automatic creation (which may contribute to the poor confusability of some of the entries), such as identical payloads. Thus, an attacker can outpace the current manual detection techniques through clever scripting. Many of the reported incidents of typosquatting with a malicious payload were active for months or even years before they were reported.

\subsection{Consequences of Typosquatting}
A package repository ecosystem consists of several distinct stakeholders, many of whom are adversely affected by typosquatting. We note some of the ways in which the consequences of confusing a perpetrator package for a victim package can be felt by these parties:

\myparagraph{Attacks against end-users}
The most subtle attack that uses typosquatting is when an adversarial uploader delivers a malicious payload as part of the dependency code, which is subsequently used as part of a user-facing application.
This attack impacts the end-user of the application. Two highly-publicized incidents of this consequence involved a malicious payload that exfiltrated sensitive information such as credit card numbers~\cite{npm_password_cvc} or cryptocurrency~\cite{npm_crypto_hijacking}.
A stealthy adversary may attempt to obscure the payload by cloning the target package and adding the malicious functionality as a Trojan.

\myparagraph{Attacks against developers using a package}
An adversary may also target the developer who mistakenly requests the perpetrator package at install time.
Both npm and PyPI allow packages to invoke shell scripts in order to configure and deploy the script, which run under the privileges of the invoking user. 
Since  packages can be installed system-wide, the user may be the administrator, opening a vector for an adversary to do catastrophic harm to the developer's machine.
A common choice for malicious package creators is to open a reverse shell, giving them full control of the victim's machine~\cite{npm_reverse_shell}.

\myparagraph{Degradation of functionality}
Even when perpetrator does not deploy malicious code, they may still hinder operations. 
If the confusion is purely accidental, it is likely to be noticed well before the victim application is deployed. 
Nevertheless, this incidental confusion will at least waste time, the victim's time, in diagnosing the problem.

\myparagraph{Latent vulnerabilities}
If a perpetrator package is not detected immediately upon installation, it may remain latent in the victim's codebase for a significant period of time.
A frequent cause of this is when a developer typosquats a target with a payload that is a clone of the current version of the package. 
While the victim experiences no initial consequences from using the wrong package, they are at the mercy of the perpetrator that the code will be kept in lockstep with the target. 
As in the case of \textit{loadsh}, mentioned in Section~\ref{sec:Introduction}, the clone may never be updated, meaning that the perpetrator is exposed to latent bugs and vulnerabilities that have been patched in the target~\cite{lodash_prototype_pollution}.

\myparagraph{Misattribution}
Even if a perpetrator package replicates all of the target functionality, it nevertheless fragments the popularity of the target package. Thus, one minor consequence of typosquatting is that the target will not get as much credit as they would without the perpetrator. Misattribution can be found in packages like \textit{asimplemde} on npm. In addition to typosquatting, this package contains identical functionality to \textit{simplemde}. References attributing credit to the original author are the sole omissions from the duplicate package.

%% file: overview.tex
\section{Detecting Typosquatting}
\label{sec:Overview}

Motivated by the number of historical instances, the ease of execution, and the severity of the possible consequences, we created \toolname{}, a tool to detect typosquatting in package repositories. At a high level, \toolname{} compares a given package name to a list of popular package names. If the given package name matches at least one of the popular packages after a set of allowed transformations (or signals), then it is considered to be a typosquatting suspect. In that case, \toolname{} raises a alert and indicates the likely package being typosquatted before prompting the user to proceed.  

\subsection{\toolname{} Workflow}

The primary way in which we expect \toolname{} to be deployed is as a user-facing utility that integrates with the package manager frontend and introspects upon packages before they are installed.

\begin{figure}[t]
\centering
\includegraphics[width=.30\textwidth]{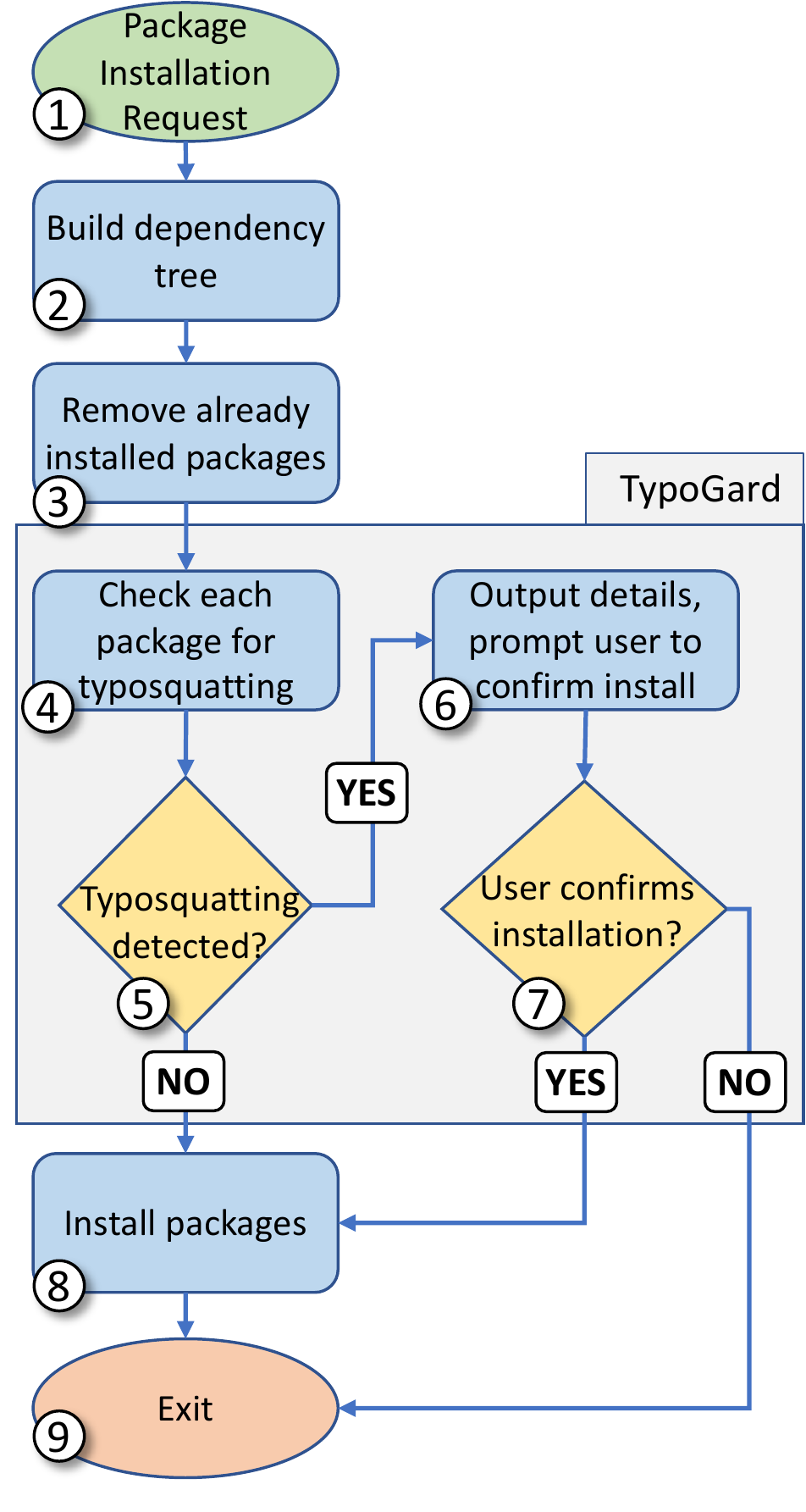}
\caption{Modified package installation process with integrated typosquatting protection.}
\label{fig:cli_modification}
\end{figure}

Figure~\ref{fig:cli_modification} depicts the overall workflow of \toolname{}, including both typosquatting detection and steps performed in the normal course of package installation. Algorithm~\ref{algo:typosquatting} presents a description of typosquatting detection (steps 4 through 7 in the figure) in pseudocode.

The user initiates the process by triggering a package's installation from the command line, e.g., \texttt{npm install loadsh} (\textbf{step 1}). The package manager computes the dependency tree of the package, i.e. its transitive closure on the package graph (\textbf{step 2}). Subsequently, it discards all packages that are already installed, and thus do not need to be downloaded (\textbf{step 3}). At this point, the workflow triggers \toolname{}'s logic.

\begin{figure}
  \includegraphics[width=0.5\textwidth]{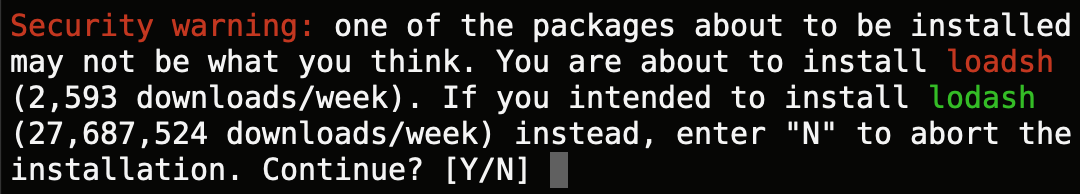}
  \caption{Package installation prompt}
  \label{fig:prompt}
  \vspace{-0.5cm}
\end{figure}

First, \toolname{} considers each package queued to be installed (\textbf{steps 4-5}, lines 1-3 in Algorithm~\ref{algo:typosquatting}). A package is considered suspicious if its \textit{popularity score} (explained in Section~\ref{sec:Popularity}) is below a tunable threshold $T_p$, and there exists a popular (popularity $\geq T_p$) package with a similar name (similarity is discussed in Section~\ref{sec:TyposquattingSignals}). If this is the case, \toolname{} flags the package and prompts the user (\textbf{step 5-6}, lines 4-8). The prompt displays a brief explanation of the warning, which includes both the name of the offending package, and the name of the package that most likely should be installed instead (an example prompt is shown in Figure~\ref{fig:prompt}). If the user decides to ignore the warning, the package is installed (\textbf{step 8}, line 6), otherwise the process is terminated. Note that \textit{AbortInstallation()} in line 8 terminates the process for all queued packages, not just the one which was the object of the warning. In lines 9-10, any package which does not raise suspicion is directly installed without prompting the user.

\subsection{\toolname{} Batch Analysis}
\label{sec:BatchAnalysis}

While we anticipate the workflow in Figure~\ref{fig:cli_modification} to be the most common application of \toolname{}, we also envision repository maintainers may want to periodically apply the same analysis in batch fashion to the \textit{entire package repository}. This would simplify the task of identifying highly suspicious packages. Our current implementation also supports this approach. In this mode, \toolname{} receives as input the list of all package names. It then returns a list of candidate perpetrators, ranked by decreasing download count. Indeed, the \textit{loadsh} package discussed in Section~\ref{sec:Introduction} was identified in this way; \toolname{}'s batch analysis ranked it as the seventh most popular typosquatting candidate matching a specific signal discussed in the next subsection.

\subsection{Typosquatting Signals}
\label{sec:TyposquattingSignals}

\toolname{} relies on the ability to identify pairs of packages with similar names; however, precisely defining the notion of similarity is challenging. Initially we experimented with thresholds on basic Levenshtein distance. However, we found this approach overly simplistic, and generating an enormous number of matches. These similarities are bound to happen purely due to the size of the repository: there are 9,371 3-letter packages in npm, and only 17,576 combinations of three lowercase English letters\footnote{Names can use other symbols, however most short names do not include them.}.

After extensively exploring alternative approaches, we designed a notion of similarity that relies on the disjunction of six possible signals (i.e., triggering one signal causes a pair of names to be considered similar). These signals were created by examining past typosquatting attacks and extending signals used to  detect domain name typosquatting such that they apply to package repositories \cite{domain_typosquatting_signals, typosquatting_homographs}. The signals, along with descriptions and examples of genuine perpetrator/victim package pairs are listed below. Note that a majority of the examples used are historical typosquatting instances and have been removed, though all examples would be detected by \tooltext.

\begin{enumerate}

    \item Repeated characters --- the presence of consecutive duplicates in
    a package name. For example, \textit{reequest} is typosquatting \textit{request}.
    \item Omitted characters --- a restricted form of edit distance, not allowing arbitrary
    character substitutions and additions. The maximum allowed
    number of omissions is set to one. For example, \textit{comander} is typosquatting
    \textit{commander} and \textit{require-port} is typosquatting \textit{requires-port}.
    
    \item Swapped characters --- two consecutive characters have been swapped.
    For example, \textit{axois} is typosquatting \textit{axios}.
    
    \item Swapped words --- this signal depends on the presence of delimiters in a
    package name, where a delimiter is a period, hyphen, or underscore. This signal
    checks for any other ordering of delimiter-separated tokens in the package repository
    namespace. This signal checks for reordering with other delimiters as well. For example, \textit{import-mysql} is typosquatting \textit{mysql-import}.
    
    \item Common typos --- character substitutions based on physical locality on the QWERTY keyboard layout. This signal also checks for substitutions of characters with visual similarity. For example, \textit{signqle} is typosquatting \textit{signale}, \textit{1odash} (with the number one) is typosquatting \textit{lodash} (with the letter L), and \textit{uglify.js} is typosquatting \textit{uglify-js}. The rationale for checking for characters with visual similarity is that, even if users are unlikely to make the typo, they may overlook such packages if they are imported indirectly as malicious dependencies. These packages are not explicitly requested by the user, however, they can be seen during the installation process. Attackers could utilize this style of substitution in hopes that it could be confused with another at a glance.
    
    \item Version numbers --- the presence of integers located at the end of package names.
    Optional delimiters between the package name and the version number are also considered.
    For example, \textit{underscore.string-2} is typosquatting \textit{underscore.string}. Note that \textit{underscore.string-2} was previously undiscovered and \tooltext led us to find a latent vulnerability.
    
\end{enumerate}

\subsection{Package Popularity}
\label{sec:Popularity}

Once the typosquatting detection scheme had been created, we required some formal definition of popularity to successfully implement \tooltext. This requirements stems from a fundamental belief that we posit, which is that only unpopular packages can be typosquatting perpetrators and only popular packages can be typosquatting targets. Popular packages are, by our definition, incapable of perpetrating typosquatting attacks. Next, we believe that there exists no incentive for an adversary to typosquat a package which receives an insignificant amount of attention. If a negligible number of users download that package, then an even smaller number of people could potentially misspell the name of that package and fall victim to the attack. By this token, a package which is downloaded thousands, millions, or even tens of millions of times per week, is a far more rewarding target.

The two main possibilities for quantifying package popularity were the number of downloads and the number of dependents. We decided to focus on the number of downloads because we believe it is a more indicative measure of true package usage. The public number of dependents counts only the number of other packages that have been uploaded to the repository that directly depend on a given package. Download count, on the other hand, counts the number of users who have downloaded that package either directly or indirectly through some arbitrarily long chain of dependencies.

Popularity based on download count requires the definition of a threshold to distinguish between popular and unpopular packages. This threshold is of crucial importance because the number of packages considered to be typosquatting depends directly on the number of packages considered to be popular. An exceedingly low threshold results in many typosquatting packages being considered popular, thus making their detection impossible. Conversely, an exceedingly high threshold may miss packages with are frequently downloaded and are victims of typosquatting. We use a data-driven approach, discussed in Section~\ref{sec:Evaluation}, to determine the threshold.

\input{algo}

%% file: algo.tex
\begin{algorithm}
  \caption{\toolname{} typosquatting detection}
  \label{algo:typosquatting}
\begin{algorithmic}[1]
  \algrenewcommand\algorithmicrequire{\textbf{Input:}}
  \algrenewcommand\algorithmicensure{\textbf{Output:}}
  \algnewcommand{\LineComment}[1]{\State \textcolor{Gray}{\textit{\(\triangleright\) #1}}}
  \algnewcommand{\InlineComment}[1]{\tabto{1.75in} \textcolor{Gray}{\textit{\(\triangleright\) #1}}}
  \algnewcommand{\Functionx}[2]{\Statex \textbf{function} #1(#2)}
  \algnewcommand\algorithmicforeach{\textbf{for each}}
  \algdef{SE}[DOWHILE]{Do}{EndDo}{\algorithmicdo}[1]{\algorithmicwhile\ #1}%
  \algdef{S}[FOR]{ForEach}[1]{\algorithmicforeach\ #1\ \algorithmicdo}
  \Require{List $I$ of packages to be installed}
  \Require{Package graph $G$}
  \Require{Popularity threshold $T_P$}
  \ForEach{$p \in I$}
    \If{Popularity($p$) $< T_P$}
      \If{$\exists p'$ s.t. Popularity($p'$) $ \geq T_P$ \textbf{and} Similar($p$, $p'$)}
        \State $R \leftarrow$ UserConfirm?($p$, $p'$);
        \If{$R = True$}
          \State Install($p$);
        \Else
          \State AbortInstallation();  
        \EndIf    
      \EndIf
    \Else
      \State Install($p$);  
    \EndIf  
  \EndFor
\end{algorithmic}
\end{algorithm}

%%% Local Variables:
%%% mode: latex
%%% TeX-master: "main"
%%% End:

%% file: evaluation.tex
\section{Analysis and Evaluation}
\label{sec:Evaluation}

In this section, we perform an in-depth analysis of \toolname{}'s tunable parameter, the popularity threshold, and we evaluate \toolname{}'s effectiveness in flagging suspicious package installs. Our goal is to answer the following questions:

\begin{enumerate}
\item Is it possible to determine an optimal popularity threshold based on repository characteristics? What is the impact of varying this threshold? (Section~\ref{sec:EvalPopularityThreshold}).
\item What is the effectiveness of \toolname{}'s typosquatting signals in identifying suspicious packages? (Section~\ref{sec:EvalSignalDetectionRates}).
\item Is the latency introduced by \toolname{} to the package installation process acceptable? (Section~\ref{sec:EvalSignalDetectionRates}).
\end{enumerate}

\subsection{Dataset}

In order to perform our analysis, we consider the entire package graphs for npm and PyPI. In particular, our analysis is based on snapshots of npm and PyPI which reflect their state on February 19, 2020. A high-level quantitative summary of both repository snapshots is given in Table~\ref{tab:repo_stats}.

\subsection{Popularity Threshold}
\label{sec:EvalPopularityThreshold}

Download counts bear an obvious relationship to the popularity of a given package within the developer community. Precisely understanding this relationship however requires careful analysis of a software ecosystem. This is due to the fact that download counts on npm and PyPI represent more than the number of people who have installed a package. Packages are regularly downloaded by repository mirrors and bots which download all packages for analysis. These downloads are also recorded in a package's total download count. Based on estimates made by the creators of npm, a package can be downloaded up to 50 times per day without ever being installed by an actual developer~\cite{npm_download_counts}.

Based on this estimate, we use 350 weekly downloads as an absolute lower bound for package popularity as packages with fewer than this number of downloads may have never been downloaded by an actual user. As seen in Figures~\ref{fig:npm_download_distribution} and ~\ref{fig:pypi_download_distribution}, a majority of packages in both npm and PyPI receive fewer than 350 weekly downloads. The stipulation that a package must have, at the very minimum, 350 weekly downloads to be considered popular removes about 93.9\% of npm packages and about 93.3\% of PyPI packages from consideration. Interestingly, this suggest that only a tiny fraction of packages in these repositories receive any meaningful attention and usage from the community. Due to the size of these repositories, however, this fraction still amounts to millions of downloads.  As an upper bound, we consider packages with more than 100,000 downloads per week to be unquestionably popular. Packages above this upper bound make up the top 0.6\% of npm and the top 0.4\% of PyPI. The analyses we describe in this section aim at finding an appropriate threshold to separate popular packages from unpopular packages between these two bounds.

\begin{figure}[t]
\includegraphics[width=\linewidth]{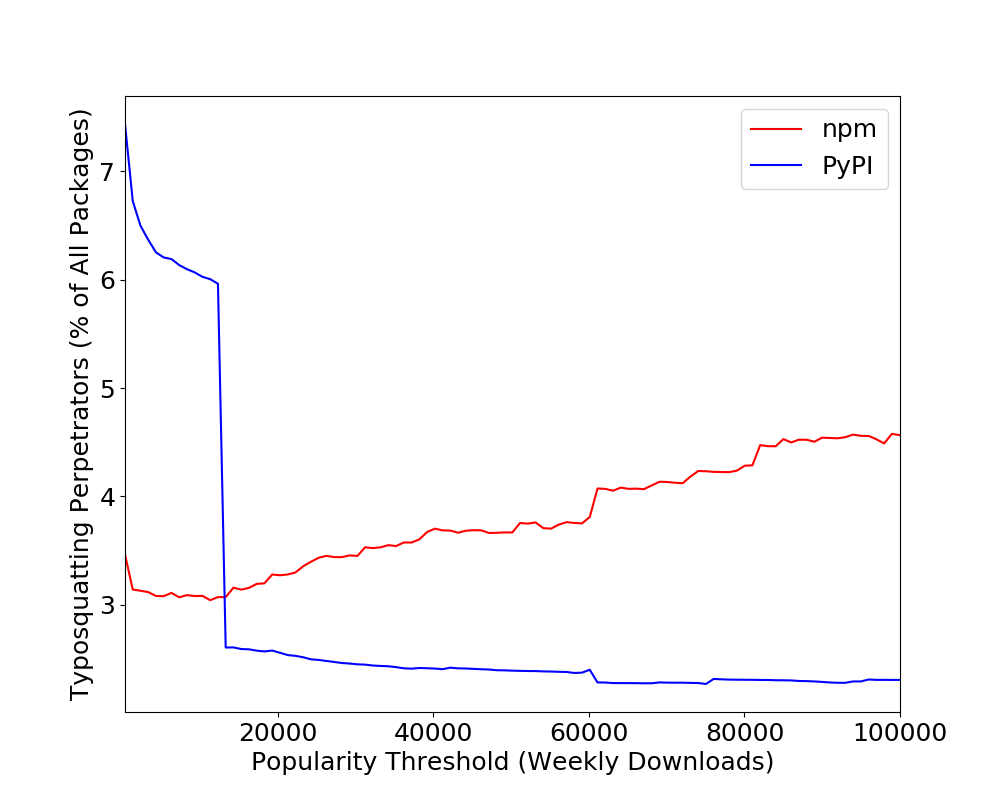}
\caption{Relationship between popularity threshold and percent of repository typosquatting.}
\label{fig:percent_repo_typosquatting}
\vspace{-0.5cm}
\end{figure}

\myparagraph{Effect of threshold on number of perpetrators} The first analysis aims to determine how the number of typosquatting targets influences the number of typosquatting perpetrators. This is a transitive test, which means a package is considered to be a typosquatting perpetrator if it, or any package in its dependency tree, fits our definition of typosquatting. Doing this emulates real-world conditions, as users typically would not install a package without installing its dependencies. The results of this analysis is depicted in Figure~\ref{fig:percent_repo_typosquatting}. Interestingly, the curves corresponding to npm and PyPI are fundamentally different. As the popularity threshold increases, the number of popular packages decreases. With this decrease in typosquatting targets, one would initially expect the number of typosquatting perpetrators to decrease. The trend for PyPI is consistent with this behavior. The sharp drop in perpetrators is due to a large number of packages that fit our definition of typosquatting that also have just over 13,000 weekly downloads. As soon as the popularity threshold crosses 13,000, these packages are considered to be popular and are therefore exempt from being typosquatting perpetrators, causing the drop in perpetrators. 

\begin{figure}[t]
\includegraphics[width=\linewidth]{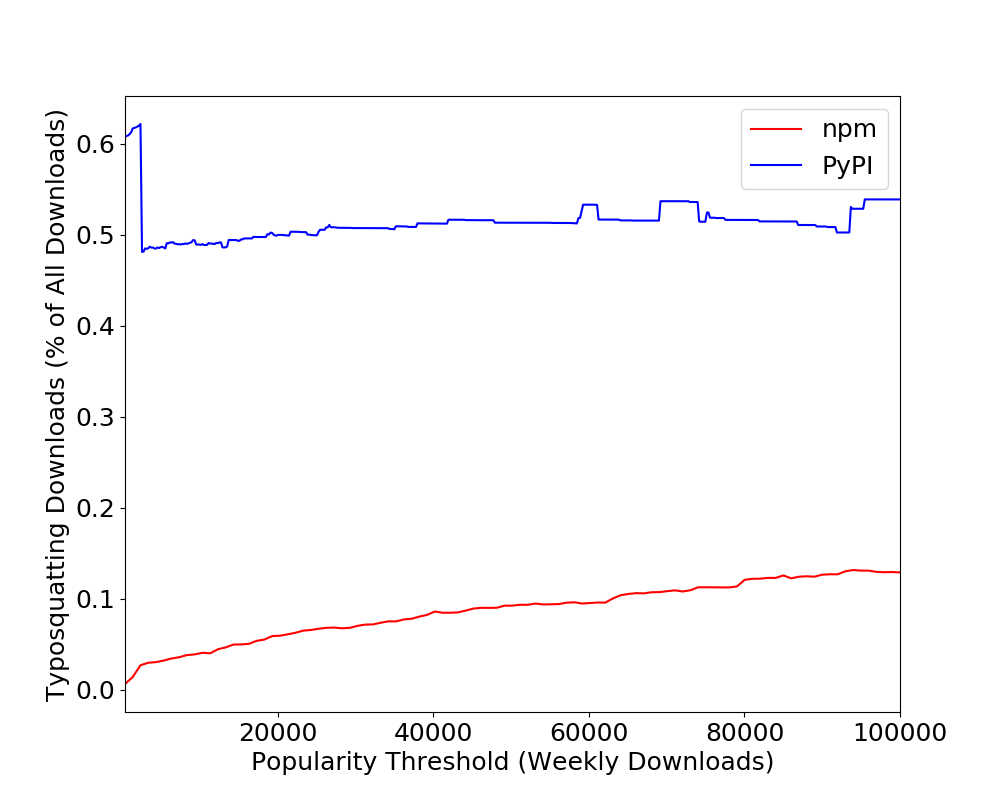}
\caption{Relationship between popularity threshold and percent of weekly downloads containing
a typosquatting package.}
\label{fig:percent_download_typosquatting}
\vspace{-0.5cm}
\end{figure}

In stark contrast contrast, npm's trend steadily increases. The number of typosquatting perpetrators grows in spite of the fact that the number of targets shrinks. This highlights an interesting phenomenon present in npm: there's a significant amount of package name similarity between reasonably popular packages. This idea is best exemplified by cases like those shown in Table~\ref{tab:popular_perps}. All of these packages have significant download counts. Examples like those found in Table~\ref{tab:popular_perps} cause the unintuitive increase in perpetrators seen in Figure~\ref{fig:percent_repo_typosquatting}. For small popularity thresholds, both packages in these pairs are considered popular. However, as the threshold grows, it passes the weekly download count of the less popular package, which in turn turns the less popular package into a perpetrator. Ultimately, this process increases the number of perpetrators as the number of targets decreases.

Based on the analysis discussed above, we have chosen to select a popularity threshold of 15,000 weekly downloads. A popularity threshold of 15,000 weekly downloads is the lowest threshold which keeps the number of typosquatting packages reasonably low for both repositories. For both npm and PyPI, approximately 3\% of all packages on each repository are potentially typosquatting for this threshold.

\myparagraph{Effect of threshold on frequency of warnings} The second analysis examines how frequently packages that could be considered typosquatting are downloaded. It is important to understand this datum in order to get a sense of how frequently \tooltext will intervene during the package installation process. Maintaining the frequency of interventions low is important for two reasons. First, frequently interrupting a developer's workflow with warning notifications risks incurring in the well-known phenomenon of warning fatigue~\cite{bohme_security_2011}. Second, it is reasonable to expect that the number of packages imported by mistake is a relatively small fraction of the overall number of packages imported by a developer. Therefore, a very high number of warning is likely to consist overwhelming of false positives~\cite{axelsson_base-rate_1999}.

This analysis, like the first, is transitive in order to emulate real-world conditions. Ideally, the number of alerts asking the user if they are sure they would like to install the requested package should be kept close to zero. The results of this analysis are show in Figure~\ref{fig:percent_download_typosquatting}. In this test, trends for both repositories are noticeably similar than the trends in Figure~\ref{fig:percent_repo_typosquatting}. According to this figure, with any reasonable popularity threshold, the percentage of weekly downloads which result in a warning from \toolname{} is around 0.1\% for npm and around 0.5\% for PyPI. In other words, \toolname{} generates on average a warning every 200 to 1000 package installs, which we consider an acceptable burden for a developer.

\begin{table}[!t]
\begin{tabular*}{.475\textwidth}{l @{\extracolsep{\fill}} r}
\toprule
Package Name  & \multicolumn{1}{l}{Weekly Downloads} \\
\toprule
object-assign & 17,249,391                           \\
object.assign & 10,843,774                           \\
\midrule
isarray       & 30,271,796                           \\
is-array      & 69,131                               \\
\midrule
is-buffer     & 19,143,770                           \\
isbuffer      & 35,684                               \\
\midrule
memorystream  & 1,125,398                            \\
memory-stream & 6,047                               
\end{tabular*}
\caption{Typosquatting cases with popular perpetrators.}
\label{tab:popular_perps}
\end{table}

\subsection{Signal Detection Rates}
\label{sec:EvalSignalDetectionRates}

In this section, we consider the effectiveness of the typosquatting signals used to determine package name similarity (ref. Section~\ref{sec:Overview}). The signals we chose to include in this implementation of \toolname{} detected approximately 60\% of known past attacks reported by the npm security team as typosquatting. While this number may appear low, it chiefly stems from qualitatively different definitions of typosquatting used by npm and us.

For example, npm considers \textit{ruffer-xor}, \textit{bwffer-xor}, \textit{bufner-xor}, and similar ones to be typosquatters of \textit{buffer-xor}. While the former names are all at a Levenshtein distance of 1 from the target package, it is unlikely that a developer would purposely import any of the former packages in place of \textit{buffer-xor}. Typos are likewise unlikely due to the significant distance between swapped character on most keyboard layouts. As elucidated in Section~\ref{sec:Overview}, we found edit distance to be a poor metric for typosquatting, and therefore we consciously avoid flagging those cases, which would result in an unmanageable number of warnings anyway.

Instead, to get a sense of how each of the signals were performing, we considered the number of packages in each repository which match a given signal. These results are shown in Table~\ref{tab:signal_counts}. Note that these figures contain no notion of dependencies and, are therefore, not transitive. Here we are interested in examining how aggressive each of the signals are. Interestingly, despite npm having about 6 times as many packages as PyPI, the number of npm packages which fit our definition of typosquatting is almost 10 times higher. This result points toward the conclusion that typosquatting is inherently a larger issue in npm.

% Maybe discuss the loadsh case here

\subsection{\tooltext Overhead}
\label{sec:EvalPerformance}

The goal of our final analysis of \toolname{} is to determine the temporal overhead it imposes on the package installation process. To quantify the performance of \toolname{}, 1,000 npm packages were selected at random, weighted by popularity. Weighting the selections during this process is crucial, as it creates a sample that simulates the downloading patterns of actual repository users. Once selected, the contents of these packages were locally cached to remove any uncontrollable network-based effects on installation times. After being cached, installation times for each package were measured using npm's official package manager and a version modified to implement \toolname{}. The official npm package manager had an average installation time of 2.604 seconds, while \toolname{} resulted in an average installation time of 2.669 seconds, meaning \toolname{} imposes an \textbf{average temporal overhead of about 2.5\%}. We believe this result is reasonable and the slowdown incurred by \tooltext is effectively unnoticeable.

\myparagraph{Batch mode performance} Batch mode (ref. Section~\ref{sec:BatchAnalysis}) analyzes the entire package set in a single pass and is intended to be used by repository maintainers to discover yet unknown issues of typosquatting. In our experiments, we found that \toolname{} can analyze the entire npm package set in \textbf{11 minutes}. This result suggests that \toolname{} could be run frequently (e.g., once per day) allowing quick identification of unknown typosquatting cases.

\begin{table}[!t]
\begin{tabular*}{.475\textwidth}{l @{\extracolsep{\fill}} ll}
\toprule
                    & npm   & PyPI \\
\toprule
Repeated Characters & 443   & 40   \\
Omitted Characters  & 3827  & 412  \\
Swapped Characters  & 514   & 63   \\
Swapped Words       & 1732  & 77   \\
Common Typos        & 4409  & 533  \\
Version Numbers     & 1148  & 116  \\
                    &       &      \\
\textbf{Total}      & \textbf{12073} & \textbf{1241}
\end{tabular*}
\caption{Number of packages triggering each typosquatting signal.}
\label{tab:signal_counts}
\end{table}

%%% Local Variables:
%%% mode: latex
%%% TeX-master: "main"
%%% End:

%% file: discussion.tex
\section{Discussion}
\label{sec:Discussion}

In this section, we discuss the broader implications of our findings, include some of the subtleties related to typosquatting, explore possible steps to mitigate typosquatting beyond \tooltext, and consider alternative ways to implement \tooltext.

\subsection{Alternative \tooltext Deployments}
As discussed in Section~\ref{sec:Overview}, our primary deployment of \tooltext is a modification to existing package manager frontend tools. Implementing our tool in this way allows typosquatting protection to be non-invasive and fit into existing workflows. Ultimately, we hope that our mechanism is incorporated into existing package management tools. However, we also implemented a standalone command-line tool that performs our transitive typosquatting protection checks without the cooperation of the frontend, thus allowing users to avail themselves of typosquatting protection even if such protection is not directly integrated in the package manager.

The goal of \tooltext is to decrease the chances that a user of a package manager will accidentally install an incorrect package due to typosquatting. 
However, it is beyond the scope of this work to model all of the ways in which a user might confuse their target package name. For example, confusion may stem from misremembering a name, or hearing it incorrectly. Similarly, the particular keyboard layout used by a developer influences the typos that that developer is likely to make when typing in the package name. Collectively, these differences may justify personalizing the typosquatting detection scheme.

%The ways in which a user might arrive at the wrong package are numerous, and often depend upon the context by which the user decides to incorporate the package: if they read about the package, they may be more likely to fall victim to a package name that is visually similar, such as installing a package that uses the numeral 1 instead of the lowercase l.
%If they hear about a particular package, they may be more likely to fall victim to a package whose name is a homonym of another.
%Similarly, alternate keyboard layouts influence which keys could be typed mistakenly when invoking a package name. Collectively, these differences may justify personalizing the typosquatting detection scheme.

\tooltext relies on the concept of popularity.
It is possible to define alternative notions of popularity by changing the metric with which the popularity of a given package is quantified (e.g. using the number of dependent packages). Exploring these alternatives is future work.
An additional implementation detail of our detection algorithm is that it considers potential victim packages and potential perpetrator packages to be disjoint sets partitioned by the popularity threshold. 
A natural extension would be to consider these sets to overlap, such that somewhat popular packages could be classified as typosquatting perpetrators or victims.

The evaluation results in Section~\ref{sec:Evaluation} show that the perpetrator package detection algorithm developed as part of this work is unobtrusive, but detects real cases of typosquatting. Nevertheless, the modular design of \tooltext means that the alternative approaches outlined above could be dropped in to the tool with no changes to the workflow. 

% this means, for npm, as the number of typosquatting targets decreases,
% the number of typosquatting perpetrators increase
% this means some potential targets could also be potential perpetrators
% mention object-assign vs object.assign

\subsection{Server-Side Protection Mechanisms}

Our technique successfully detected typosquatting that was active in popular package repositories for over a year, leading to effective remediation: developers updated their dependencies to their intended target package, and repository maintainers seized and deprecated the perpetrator package. 
Consequently, we feel that our approach could aid server-side security teams in scanning their entire repository to discovered latent typosquatting instances.
As discussed in Section~\ref{sec:Overview}, repository maintainers can run \tooltext in batch mode to identify suspicious packages that have already been uploaded. We also consider some additional mechanisms that may help to combat the typosquatting problems.

\myparagraph{Preemptive takedown}
An aggressive extension to server-side batch mode operation of \tooltext is to invoke a typoquatting check at the time a new package is uploaded, effectively disallowing the existence of too-similar package names.
This proactive approach is a natural extension to the case-insensitive and delimiter-based naming restrictions currently in place on npm and PyPI~\cite{npm_monikers, npm_crossenv, pypi_naming}. It further limits the potential of a perpetrator package from gaining traction and achieving legitimacy through the confusion of users.
We note that an implicit assumption of our current approach is that popular packages cannot, by definition, perpetrate typosquatting attacks.
Our definition means that if an illegitimate package gains enough traction to exceed the threshold, it can avoid triggering a warning on installation. Disallowing the perpetrator package from being uploaded obviates that issue. 

\myparagraph{Variant-insensitive package names}
Much like disallowing too-similar package names, a repository could map all variations of a package name to the canonical version of the package. This approach means that the perpetrator would be unable to upload their package, since the system would consider the name to be taken by the target package. Furthermore, it would address the typo by suggesting the correct target.
Some repositories already implement some limited form of this behavior. PyPI maps all punctuation to hyphens and handles all package installation requests in a case-insensitive manner~\cite{pypi_naming}.
We believe such changes warrant future research. 
A potential concern with allowing all variations of a name to map to the same package is that it crowds the set of possible names. We note that npm already incorporates a typo-safe mechanism to allow similar package names, called \textit{scoped packages}~\cite{npm_scope}. The mechanism works by allowing package names to begin with an @ symbol, followed by a namespace portion (typically the package creator's username), followed by a forward slash, followed by the basename of the package. Versions of many popular packages deployed using TypeScript (a typed superset of JavaScript) are available under the @types/ namespace (e.g. @types/node for the TypeScript version of the node package). Scoped packages can be used to alleviate the concern that a repository's names may become too crowded for a new package to be given a descriptive name. 

\subsection{Defensive Typosquatting}

One tactic currently used to prevent package typosquatting is to preemptively register confusable variants alongside the canonical package name, so that the variants cannot fall under the control of a typosquatter. 
We refer to this tactic as \textit{defensive typosquatting}. 
In the absence of officially supported mechanisms for defensive typosquatting, a benign placeholder package will be registered under the variant name.  
We observed instances of defensive typosquatting in both npm and PyPI, where package creators (or 3rd party package developers) are free to create as many packages as they desire with varying behaviors for placeholder packages. The placeholder behaviors that we observed are as follows:

\myparagraph{Transparent inclusion of target package functionality}
One approach is to transparently provide the functionality of the target package to the user within the placeholder package. 
This approach can be accomplishing with varying degrees of sophistication.
By leveraging the repository's dependence mechanism, a placeholder package creators can effectively implement a passthrough to the intended target package. By making the legitimate package a dependency of the placeholder, the correct package is installed despite the request being incorrect. When included in a project, the defensive typosquatting package can simply import the legitimate package's functionality. 
We observed this behavior in practice in the npm package \textit{buynan}, which (defensively) typosquats the legitimate package \textit{bunyan}. The \textit{buynan} package simply imports \textit{bunyan} upon its inclusion.
One limitation of this defense is that it is indiscernible from a case of a malicious Trojan package; at any point a 3rd-party owner of a placeholder could change the redirect to a malicious payload. 
Furthermore, a less sophisticated method for transparently including target package functionality is to clone the code of the target. However, if the placeholder fails to stay up-to-date with the package it defends, it can actually expose the user to latent vulnerabilities, effectively becoming a stale package typosquatting perpetrator. This was the particular situation \textit{loadsh} was in when discovered. 

\myparagraph{User alerts}
One possible option is to make the placeholder issue an informative alert with directions to change to the legitimate package. 
For example, the placeholder could print a message during at install time or runtime to inform the users of their mistake. This approach has been extensively used within the PyPI repository~\cite{bommarito2019empirical, pypi_parking_1}. In this case, placeholder packages utilize the install hook mechanism of PyPI to issue a message at install time that directs users to the packages they likely had in mind.

\myparagraph{Package Deprecation}
One mechanism used in practice to alert users that they should change packages is the deprecation mechanism. This mechanism allows a package maintainer to indicate that it should no longer be used. When a deprecated package is installed, the user is presented with an alert.
Deprecation is used in practice when a stale package typosquatting perpetrator is discovered, since it does not break dependant code but still admonishes victims to update their dependencies. One limitation of this technique is that deprecation is a mechanism that is used for a variety of purposes. It is unclear whether the deprecation mechanism is sufficient to alert users of the type of error that they have incurred.

Defensive typosquatting will continue to have a place as a stopgap mechanism to protect against package name confusion, even in the presence of automated defenses like \tooltext, since the confusion may occur due to the specific context of the package name. Nevertheless, tools like \tooltext can ease the burden of placing placeholder packages.
%%% Local Variables:
%%% mode: latex
%%% TeX-master: t
%%% End:

%% file: related_work.tex
\section{Related Work}
\label{sec:related_work}

\myparagraph{Typosquatting and Defenses}
Tschacher's Bachelor thesis~\cite{tschacher_typosquatting_2016} demonstrates high success rates of a controlled typosquatting attack, proving the importance of devising countermeasures. It also briefly outlines defenses based on forbidding names similar to those of popular packages, but does not implement or evaluate them, and does not consider involving developers in the decision process. 

The creators of npm and PyPI have taken basic countermeasures to combat typosquatting. Rules governing package names for these platforms have become increasingly strict in an attempt to mitigate the problem. Both platforms have incorporated restrictions on capitalization and punctuation-based differences~\cite{npm_monikers, pypi_naming}. User-led defense campaigns exist that aim to "park" potential typosquatting names before they can be used in a malicious manner~\cite{pypi_parking_1, pypi_parking_2}.

\myparagraph{Domain Name Typosquatting}
Domain name typosquatting has long been a popular attack vector, allowing cybercriminals to hijack web communications~\cite{spaulding_landscape_2016} and potentially emails~\cite{szurdi_email_2017}. Such hijacking is typically operated by registering a domain name similar to a popular one. It is used for financial gain, by serving ads, pushing drive-by downloads, or orchestrating phishing attacks. In particularly serious cases, regulations such as the US ACPA allow ICANN to seize typosquatted domains to prevent confusion~\cite{acpa1999}. Such legal framework does not exist for package typosquatting, and indeed this approach may be difficult to apply due to the fast-evolving nature of software ecosystems. Furthermore, not all instances of package name typosquatting are the result of explicit attacks.

\myparagraph{Software Ecosystem Security}
Most past efforts focused on vulnerabilities of package managers themselves~\cite{cappos_look_2008, anish_athalye_package_nodate}, or potential attack strategies enacted by malicious packages~\cite{Pfretzschner:2017:IDA:3098954.3120928}. Both goals are orthogonal to ours, and none of these works reviewed actual incidents or performed measurements on the extent of the problem.

Other works more specifically analyze security risks arising from the presence of malicious packages in highly interconnected software ecosystems~\cite{hejderup_dependencies_2015,zimmermann_small_2019}. \cite{zimmermann_small_2019} also identifies typosquatting as one of multiple possible avenues for attack, but it provides no in-depth analysis of the phenomenon, nor describes solutions.

%Finally, Tschacher's Bachelor thesis~\cite{tschacher_typosquatting_2016} demonstrates high success rates of a controlled typosquatting attack, proving the importance of devising countermeasures. It also briefly outlines defenses based on forbidding names similar to those of popular packages, but does not implement or evaluate them, and does not consider involving developers in the decision process.

\myparagraph{General Characterization of Software Ecosystems}
Literature presents many other analyses of software ecosystems. While
these works present useful information for understanding these complex objects, they do not focus on typosquatting or other potential security-related issues. Examples include~\cite{german2013evolution, raemaekers2013maven, wittern2016look}. 

\myparagraph{Mobile Ecosystems}
A related line of work is on the study of mobile application markets such as
the Google Play store~\cite{viennot2014measurement,chatterjee18-ipv, wermke18,
chakradeo13}. These works are primarily concerned with
applications used by consumers, rather than application components (packages) that are specific to the language ecosystem and used by developers. As such, characterization of app markets (and
defenses proposed against malicious applications) are largely
orthogonal to our work. The closest work is in the 
detection of \textit{cloned} applications, whereby a
lesser-known or actively malicious developer will re-package and
re-publish a better-known app. Detecting application clones has
typically been done via code similarity metrics~\cite{droidkin}
or behavior~\cite{andarwin}. In contrast, our approach is based entirely on the package metadata and an analysis of the properties of the package repository.

\myparagraph{Supply Chain Vulnerabilities}
Others have looked at the related problem of \textit{supply chain vulnerabilities}, i.e., vulnerabilities in the open-source applications on which a software package depends~\cite{tellnes_dependencies:_2013,cadariu_tracking_2015,younis_using_2014,plate_impact_2015,kula_visualizing_2014}. These works typically discuss identification or impact of potential upstream vulnerabilities. While an attacker could attempt to introduce such a vulnerability via typosquatting, analyzing this possibility is outside the scope of our work.

%%% Local Variables:
%%% mode: latex
%%% TeX-master: "main"
%%% End:

%% file: conclusion.tex
\section{Conclusion}
\label{sec:Conclusion}

Package managers vastly improve the software development workflow. They can
quickly download and install third-party packages, along with any dependencies,
to import constructive functionality into a project. Packages are typically
requested explicitly by name and currently, there exists no safety net for
developers during the package installation process. Typosquatting attacks
target those who make a spelling mistake and their effects can be severe.
These attacks are far from novel due to their extensive history of targeting
domain names. Although, the focus of typosquatting attacks has recently grown to
include package repositories. Despite hundreds of past attacks, practical
defenses against typosquatting in package repositories such as
npm and PyPI have received little attention.

In this paper, we have shown that a defense against these attacks is both
practical and efficient. By comparing the name in the requested package's
dependency tree to a list of probable targets, our proposed solution
can protect developers from typosquatting attacks. With an average overhead
of 2.5\%, a warning-to-install ratio of 0.5\%, and third-party confirmation of flagged packages, our solution imposes
a negligible burden while protecting package creators and end users alike.

% \myparagraph{Data Policy} All data is available for review at {\it \url{https://www.dropbox.com/sh/wrkz2l3njol0ecw/AAAqbv9hN83Cfdq2CGy6bBjma?dl=0}}. After publication, the same data will be made publicly available on our institutions' platforms for long-term data publishing.

%%% Local Variables:
%%% mode: latex
%%% TeX-master: t
%%% End: